\documentclass[12pt]{article}

\usepackage{epsfig}
\usepackage{latexsym}
\usepackage{amsmath}
\usepackage{amssymb}
\usepackage{psfrag}
\usepackage{citesort}
\usepackage{graphicx}

\begin{document}

\begin{center}
{\LARGE{\bf Variations in Structure Explain the 
Viscometric Behavior of AOT Microemulsions
at Low Water/AOT Molar Ratios}}
\\

{\Large S. Sharifi$^1$, P. Kudla$^2$, 
C.~L.~P. Oliveira$^3$, J.~S. Pedersen$^4$, 
and J. Bergenholtz$^{5,*}$}
\\

{\em $^1$ Department of Physics, University of Sistan and Baluchestan,
98135-674 Zahedan, Iran; $^2$ Structure analysis, Beiersdorf AG,
D-20245, Hamburg, Germany; $^3$ Instituto de F\'{i}sica, 
Universidade de S\~ao Paulo, Caixa Postal 66318, 05314-970,
S\~ao Paulo, Brasil; $^4$ 
Department of Chemistry and iNANO Interdisciplinary
Nanoscience Center, University of Aarhus, DK-8000, Aarhus C,
Denmark; $^5$ Department of Chemistry, University of Gothenburg,
SE-41296 G\"oteborg, Sweden
}
\end{center}

\noindent The viscosity of AOT/water/decane water-in-oil microemulsions
exhibits a well-known maximum as a function of water/AOT
molar ratio, which is usually attributed to increased
attractions among nearly spherical droplets. The
maximum can be removed by adding salt or by changing
the oil to CCl$_4$. Systematic small-angle X-ray scattering
(SAXS) measurements have been used to monitor the structure
of the microemulsion droplets in the composition regime
where the maximum appears.
On increasing the droplet
concentration, the scattering intensity is found to scale with
the inverse of the wavevector, a behavior which is consistent with
cylindrical structures.
The inverse wavevector scaling is not observed when the molar ratio is
changed, moving the system away from the value corresponding to the viscosity maximum.
It is also not present in the scattering from systems containing enough added salt
to essentially eliminate the viscosity maximum.
An asymptotic analysis of the SAXS data, complemented by
some quantitative modeling,
is consistent with cylindrical growth of droplets as their concentration
is increased. Such elongated
structures are familiar from related AOT systems
in which the sodium counterion has been exchanged for
a divalent one. However, the results of this study suggest
that the formation of non-spherical aggregates
at low molar ratios is an intrinsic property of AOT.

\section{Introduction}

The anionic surfactant Aerosol OT (sodium bis-(2-ethylhexyl) sulfosuccinate, AOT)
together with water and an oil readily forms ternary microemulsions without cosurfactants.
Such AOT-based systems form a variety of self-assembled structures in solution;
in particular, they exhibit a generous
single-phase, water-in-oil (L$_2$) microemulsion
region in terms of composition and temperature in which
the structure is rather simple and characterized by
water droplets in oil with an AOT-stabilized interface.
As is well known, the size of the droplets can be controlled
by the water-to-AOT molar ratio $X$
and the oil concentration provides control
of the droplet concentration \cite{Zul79,Day79}.
L$_2$ phases of AOT and water with liquid alkanes have
been the subject of numerous studies \cite{De95,Mou98}.
Viscosity measurements have revealed
a maximum at low-to-moderate molar ratios as the
droplets are gradually swollen by adding water at constant
droplet mass fraction \cite{Rou79a,Rou79b,Pey88,Ber95,Bat99}.
Given that the measured droplet size appears to vary
linearly with molar ratio
\cite{Zul79,Day79} and that interactions are
expected to do the same \cite{Hua85},
one expects a monotonically increasing viscosity instead of a maximum \cite{Ber95,Bat99}.
In addition to the viscosity, the osmotic compressibility and
scaled collective diffusion coefficient
exhibit corresponding minima as a function of molar ratio, which
persist even on diluting the microemulsions down to concentrations
where only single-particle properties and pair interactions
matter \cite{Bat99}. Since the relatively low intrinsic viscosities measured
suggested close to spherical droplets, the origin of the maximum was
attributed to increased attractive interactions
between droplets that are close to spherical in shape
\cite{Ber95,Bat99,Hir95,Pan08}.

Several other properties are known to change irregularly
at these lower molar ratios.
The apparent molar volume of the solubilized
water deviates significantly from the bulk value \cite{Ama00,Yos00}
and the apparent AOT head group area varies strongly with
molar ratio in this composition region \cite{Eic76,Mai84}.
Water motion is restricted \cite{Won77} and interfacial and core
water fractions can be identified
at molar ratios near the viscosity maximum \cite{Zin79,Dok06}.
The conclusion that attractions cause the viscosity maximum
as well as the associated minima in the second virial
coefficient and the dilute collective diffusivity rests
on droplets being spherical in shape.
The modest intrinsic viscosities measured
do not deviate much from the sphere value \cite{Eas93b,Ber95,Bat99}
and have consequently given little cause for concern in this regard.
However, in terms of more detailed structural characterization,
the composition regime where the viscosity maximum is observed
has been mostly overlooked as previous small-angle
X-ray and neutron scattering
studies have focused primarily on micelles \cite{Kot85}
and swollen systems at higher molar ratios \cite{Kot84,Rob89,Arl01}.
More recent studies using self-diffusion NMR
\cite{Pit00} and computer simulations \cite{Gar07} do give some
cause for concern as according to them non-spherical droplet shapes seem to be
the rule rather than the exception at low molar ratios.
Moreover, microemulsions with cylindrical droplet structures are predicted
from mean-field theory for
systems with an actual radius of curvature different from the preferred one \cite{Saf84}.

Similar viscosity maxima, although far more extreme,
are observed in lecithin/oil systems upon addition of small amounts of water, where
these systems produce highly viscoelastic gels in a narrow interval of water-lecithin molar ratios
ranging between 1 and 12 depending on the continuous phase \cite{Lui90}.
In contrast, at similar molar ratios the AOT systems remain viscous, Newtonian liquids
and exhibit a modest viscosity enhancement at the peak.
Nevertheless, given that the viscoelasticity and the viscosity maxima of the lecithin systems
are caused by a water-induced structural transformation to polymer-like aggregates
\cite{Lui90,Sch91}, it seems an examination of the structure of the AOT
microemulsions in the composition regime of the viscosity maximum is warranted.

It is against this background that we set out to perform a
systematic structural study of AOT/water/decane
microemulsions using small-angle X-ray scattering (SAXS)
at compositions corresponding to those where the viscosity maximum
is observed.  The purpose is to examine whether the viscosity
maximum is caused by increased attractions among nearly spherical
microemulsion droplets,
as concluded in the past \cite{Ber95,Bat99,Hir95,Pan08},
or whether structural changes need to be accounted for.
To this end, we seek for ways to reduce or remove
the viscosity maximum. Exchanging the alkane continuous phase
for CCl$_4$ accomplishes this and so does addition of sufficient amounts
of salt. Here we have used tetrabutyl ammonium chloride (TBAC),
where TBA$^{+}$ has a 20-fold higher affinity for the AOT
molecule than does the usual Na$^+$ counterion \cite{Hav00}.
Comparison between microemulsions with and without TBAC,
leading to microemulsions with and without a viscosity maximum,
allows for identifying structural features in the SAXS spectra
tied to the viscosity maximum. In short, the data are consistent with
droplets undergoing gradual
cylinder-sphere transitions with increasing $X$ and cylindrical
rather than globular shapes being favored with increasing
droplet concentration. Considered in a broader context,
the results suggest that such shape changes,
at least qualitatively, are an intrinsic property of the AOT molecule
rather than being intimately connected to the choice of counterion
\cite{Pet91,Eas92,Eas93a,Eas93b,Eas94}.

\section{Experimental}

Sodium bis-(2-ethylhexyl) sulfosuccinate
($\ge $99\% purity), tetrabutylammonium chloride,
and {\em n}-decane were obtained from Sigma-Aldrich
and the carbon tetrachloride used was from Riedel-de Ha\"en.
All chemicals were used as received and MilliQ water was used
in preparing all samples.
Microemulsions were prepared by weight, in terms
of the mass fraction of droplets (AOT and water)
and the molar ratio $X$=[H$_2$O]/[AOT],
and all experiments were done at a temperature of 20$^\circ$C.
In cases of salt-containing microemulsions, aqueous salt solutions
were used in place of water. Samples were
filtered using 0.2-$\mu $m Teflon filters (Gelman).

Density measurements were made
using a precision density meter (DMA 5000, Anton Paar) on
oil dilution series of microemulsions at constant H$_2$O/AOT
and TBAC/AOT molar ratios. The reciprocal sample density
behaved in all cases as a linear function of the
droplet (AOT+H$_2$O+TBAC) mass fraction,
which indicates that the mixture of droplets and solvent behave as an ideal mixture.
This was used to extract a droplet density, which
was used to determine sample densities required for
converting the kinematic viscosity from the viscometry
measurements to an absolute viscosity as well as to convert
droplet mass fractions to droplet volume fractions.
Viscosity measurements were carried out using
three dilution-type Ubbelohde capillary viscometers (Cannon Instrument Co.),
covering roughly the viscosity range 1.6-35 mPa$\cdot $s.
The temperature of the viscosity measurements was controlled
to within $\pm 0.03^\circ$C by keeping the
viscometers submerged in a constant-temperature water bath.
Several solutions were examined with rotational
rheometry (Paar Physica MCR300)
using a double-gap concentric cylinder measuring system
in controlled strain mode.
Without exception they behaved as Newtonian liquids
over an extended range of shear rates, confirming
the findings of Peyrelasse et al. \cite{Pey88}.

Small-angle X-ray scattering (SAXS) measurements were performed using
the pinhole SAXS instrument at the University of Aarhus \cite{Ped04}.
The instrument consists of an X-ray camera (NanoSTAR, Bruker AXS)
with a rotating anode X-ray (Cu K$_\alpha $ radiation) source,
cross-coupled G\"obel mirrors, collimation using three pinholes
and an evacuated beam path, and a 2D position-sensitive gas
detector (HiSTAR).
In the current experiments pinholes were used to
give a range of scattering vectors as $0.0084<q($\AA$^{-1})<0.35$,
where $q=(4\pi/\lambda )\sin(\theta /2)$ is the modulus of the
wave vector, $\theta $ is the scattering angle,
and $\lambda =1.542 $\AA\, is the X-ray wavelength.
Samples were held in 1 mm quartz capillaries, which were thermostated
to a temperature of 20$^\circ $C.
Scattering data were corrected for detector
efficiency and for spatial distortions, and azimuthally averaged.
The scattering from {\em n}-decane in the same capillaries
was measured as background and was subtracted to yield
the excess scattering as a function of $q$ for microemulsion samples.
In addition, data were converted to an absolute intensity scale
by using water as a standard.

\vskip0pt plus2mm
\section{Results and discussion}

\vskip0pt plus2mm
\subsection{Viscometry}

Building on early studies by Rouviere et al. \cite{Rou79a,Rou79b},
Peyrelasse and co-workers showed that for a range of
hydrocarbon oils the viscosity of AOT w/o microemulsions exhibits
a maximum as a function of water/AOT molar ratio $X$
at constant droplet concentration \cite{Pey88}.
The maximum occurred at $X\sim 5-8$, which is a region
where the behavior of the entrained water deviates strongly from that of
bulk water. In other words, there is some question as to whether the
system should be thought of as a micellar system with
bound water or a microemulsion system with solubilized water.
Moreover, they observed a strong correlation between the
viscosity maximum and the conductivity percolation
threshold \cite{Pey88}, something that was rationalized
later in terms of percolation
theories of attractive spheres \cite{Pey93}.

\begin{figure}[ht]
\centerline{\psfig{figure=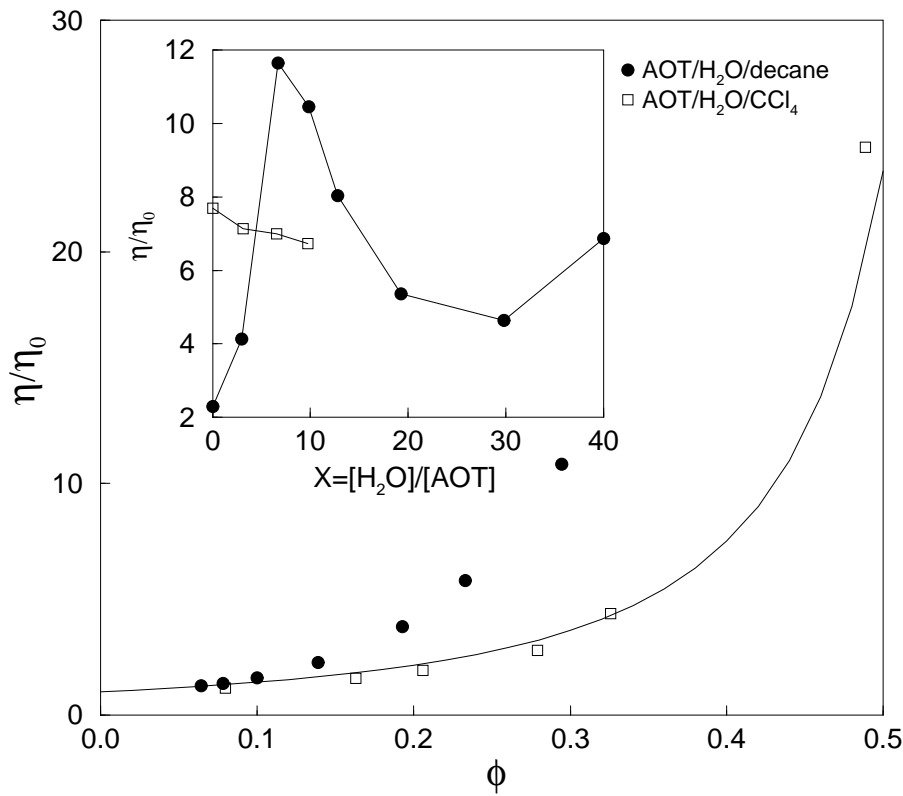,width=12 cm}}
\caption{Relative viscosity as a function of
droplet volume fraction for AOT/H$_2$O/decane
and AOT/H$_2$O/CCl$_4$ microemulsions
at a H$_2$O/AOT molar ratio of $X=6.7$, where $\eta _0$
is the viscosity of decane or CCl$_4$.
The solid line in the main panel is the Quemada equation \cite{Que77},
given by $\eta /\eta _0 =(1-\phi /0.63)^{-2}$.
The inset shows the relative viscosity as a function
molar ratio for AOT/H$_2$O/decane and AOT/H$_2$O/CCl$_4$
microemulsions at a constant droplet mass fraction of 0.3.
Note that the droplet volume fractions differ
between the decane and CCl$_4$-based microemulsions in the inset.
}
\label{ccl4}
\end{figure}

In the study by Peyrelasse et al. \cite{Pey88}, the choice of
oil had little effect on the viscosity.
However, past dilute viscosity measurements \cite{Ono92,Ber95}
suggest that using CCl$_4$ as continuous phase should
have a strong effect on the viscosity of non-dilute AOT microemulsions.
Figure \ref{ccl4} shows the relative viscosity as a function of
droplet concentration for microemulsion droplets in decane and CCl$_4$
at the same molar ratio, [H$_2$O]/[AOT]=6.7.
As seen, the viscosity increases far more rapidly with droplet
volume fraction for AOT/H$_2$O/decane microemulsions
than for AOT/H$_2$O/CCl$_4$ microemulsions.
In fact, the relative viscosity of AOT/H$_2$O/CCl$_4$ microemulsions
agrees quite well with the Quemada equation \cite{Que77},
$\eta /\eta _0 =(1-\phi /0.63)^{-2}$,
which captures rather well the viscosity of dispersions of
somewhat polydisperse
hard-sphere particles \cite{deK86}.
This difference in viscosity of the two systems
persists along the oil-dilution line in the phase diagram
down to very low droplet concentrations \cite{Ber95}.
Similar measurements at varying molar ratio produce
the data shown in the inset to Fig.~\ref{ccl4}.
As seen, a pronounced maximum in
viscosity appears for intermediate
values of the molar ratio, centered around $X\approx 6.7$,
which has been the subject of
previous study \cite{Rou79a,Rou79b,Pey88,Ber95,Bat99,Hir95,Pan08}.
As also seen, the maximum is removed on exchanging decane for
CCl$_4$, consistent with measurements under dilute
conditions \cite{Ber95}. In the AOT/water/decane case the
maximum is followed
by another increase in viscosity at higher molar ratios, which
is not observed for, e.g., AOT/water/heptane microemulsions
at the same temperature.
This increase is connected to the proximity of the lower
critical solution curve \cite{Kot84}, which is approached by
increasing the molar ratio at constant volume fraction of droplets.
Similar maxima at low molar ratios
are reported for a range of liquid alkanes \cite{Pey88}.
However, with CCl$_4$ as the continuous phase
the maximum is eliminated, showing
that solvent penetration in the surfactant tail region
is an important factor for the observed behavior \cite{Che86b}.

\begin{figure}[ht]
\centerline{\psfig{figure=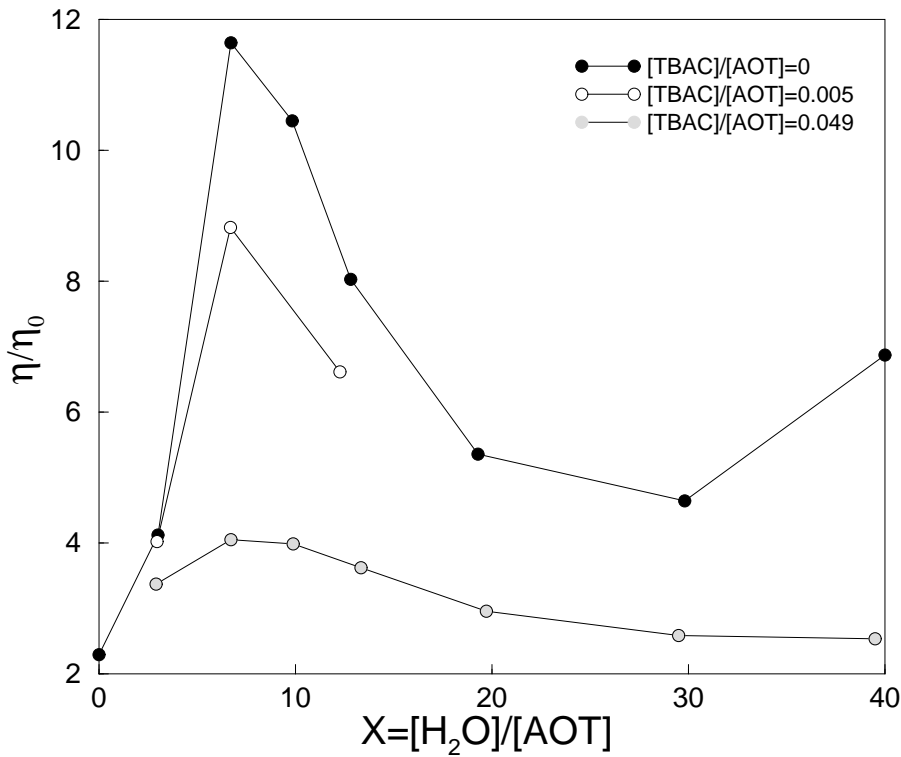,width=12 cm}}
\caption{Relative viscosity as a function of molar ratio $X$.
The data have been measured at
a constant droplet (AOT+H$_2$O+TBAC) mass fraction of 0.3,
corresponding to a droplet volume fraction of about 0.25,
and as a function of TBAC/AOT molar ratio.
}
\label{salt}
\end{figure}

Salt has a profound effect on the viscosity in these systems,
which was reported on by Rouviere et al. \cite{Rou79b}.
The tetrabutyl ammonium cation TBA$^{+}$ of the chloride salt (TBAC)
has been found to exhibit about a 20-fold higher affinity for AOT
molecules than the usual Na$^+$ counterion \cite{Hav00}
and similar salts are known to have a strong
effect on water solubilization in AOT microemulsions \cite{Der91}.
To perturb the structure and interactions
in the system, we add varying amounts of TBAC.
As seen in Fig.~\ref{salt}, where we show the relative viscosity
of AOT/H$_2$O/decane microemulsions at a droplet mass fraction of 0.3,
addition of TBAC results in a reduction of
the viscosity maximum. Little is left of the maximum on reaching TBAC/AOT ratios
of about 0.05, which corresponds to a TBAC concentration in the
aqueous pools of the droplets of roughly 0.4 M.
Note also that as TBAC is added the viscosity maximum is reduced
in magnitude without shifting in position to other H$_2$O/AOT molar ratios.

\vskip0pt plus2mm
\subsection{Small-angle X-ray scattering results}

The viscosity maximum is eliminated either by using CCl$_4$ as a continuous
phase or by dissolving TBAC in the droplet interiors.
Presumably both alterations lead to different aggregate structures and/or
changes in inter-aggregate interactions and with
SAXS measurements we expect to
be able to distinguish between these two possibilities.
Unfortunately samples rich in chlorine, like the AOT/H$_2$O/CCl$_4$
microemulsions, cannot be studied by SAXS because of strong absorption.
Therefore, the systematic SAXS study undertaken here
has been made on decane-based
microemulsions,
varying both in terms of molar ratio and droplet volume fraction.
The study included molar ratios around the viscosity maximum,
$X= 3$, 6.8, 9, and 12.2.
In addition, we examine also
TBAC-containing samples at $X=6.75$, which, as seen in Fig.~\ref{salt},
exhibit only a very weak viscosity enhancement.

\begin{figure}[ht]
\centerline{\psfig{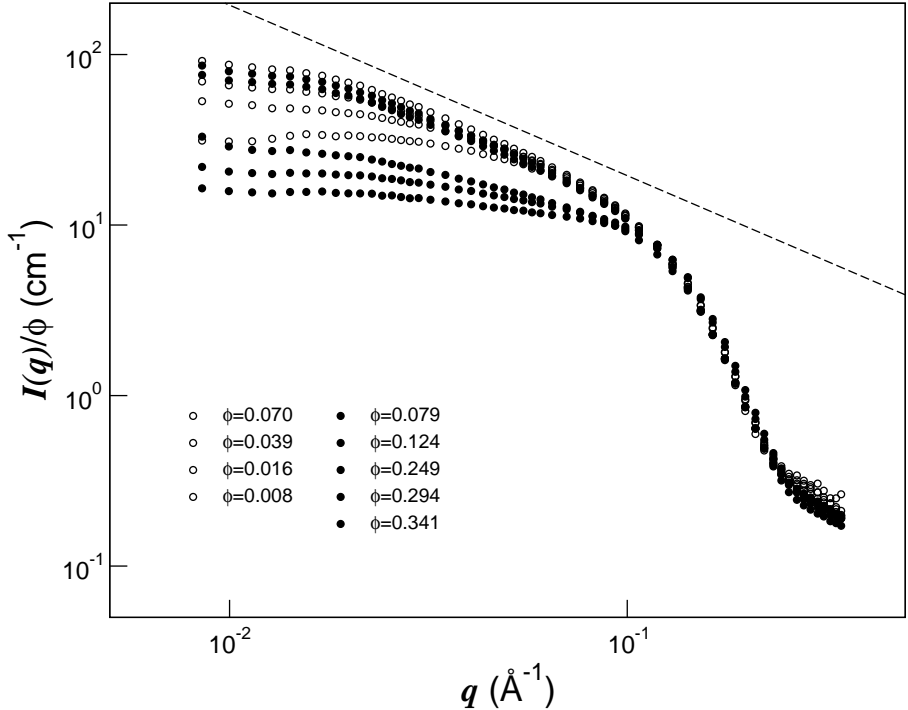}}
\caption{Intensity scaled on the droplet volume fraction $\phi $
as a function of $q$ for different $\phi $ at $X=6.8$.
The data have been divided in two groups with open circles denoting,
in order of increasing $I(q)/\phi $ along the left-hand side,
$\phi =$0.008, 0.016, 0.039, and 0.070, and
filled circles denoting, in order of decreasing $I(q)/\phi $
along the left-hand side, $\phi =$0.079, 0.124, 0.249, 0.294, and 0.341.
In other words, $I(q)/\phi $ reaches a maximum as a function of
$\phi $ at around $\phi = 0.070$.
The dashed line shows $I(q)\propto q^{-1}$.
}
\label{x6}
\end{figure}

\begin{figure}[ht]
\centerline{\psfig{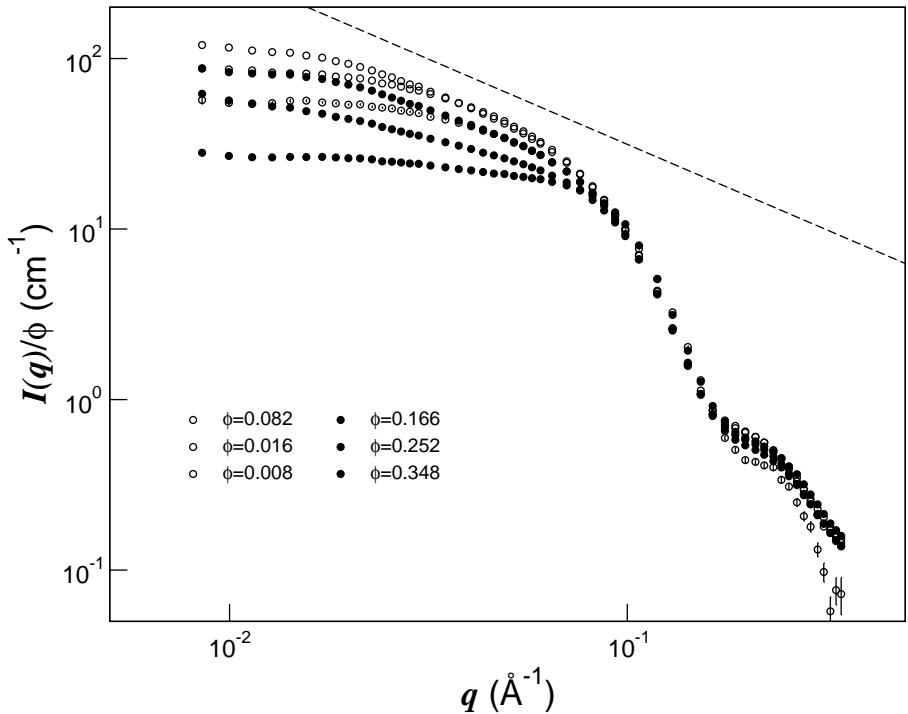}}
\caption{Intensity scaled on the droplet volume fraction $\phi $
as a function of $q$ for different $\phi $ at $X=12.2$.
The data have been divided in two groups with open circles denoting,
in order of increasing $I(q)/\phi $ along the left-hand side,
$\phi =$0.008, 0.016, 0.082, and
filled circles denoting, in order of decreasing $I(q)/\phi $
along the left-hand side, $\phi =$0.166, 0.252, and 0.348.
The dashed line shows $I(q)\propto q^{-1}$.}
\label{x12}
\end{figure}

In Fig.~\ref{x6} SAXS results are shown for salt-free
microemulsions at $X=6.8$.
This corresponds to the molar ratio of the viscosity maximum in
Figs.~\ref{ccl4} and \ref{salt}.
The intensity as a function of wavevector
has been scaled by the droplet volume fraction,
which is applied to account
for the expected leading-order, linear increase of the intensity with
the concentration of scatterers.
The figure shows that the scaled intensity initially grows with concentration.
Such an increase points to changes in aggregate structure
with concentration or possibly a negative second virial coefficient.
Since the scaling results only in a fair
collapse of the data at high $q$, where only form factor
effects are at play, the former seems more likely.
However, if so, the bend in the intensity at the highest $q$,
being positioned roughly at the same $q$ independent of concentration,
suggests that some length scale is preserved in the process
and that the structural changes occur mostly
in one or two dimensions.
The increase in $I(q)/\phi $ continues
until a regime of power-law scattering, $q^{-d}$ with
$d\approx 1$, develops at
intermediate wavevectors, a telltale sign of
linear or locally linear aggregates \cite{Por82}.
It sets in at a crossover or overlap volume fraction $\phi ^*$
of about 7\% and at this point the trend also reverses and the
scaled intensity begins to decrease gradually with increasing concentration.
The region of power-law scattering
remains beyond $\phi ^*$,
but it is now characterized by a smaller exponent.
Turning to samples of higher molar ratios, corresponding to
compositions where the viscosity has dropped relative to the
maximum value in Fig.~\ref{ccl4},
qualitatively similar behavior is observed.
Taking $X=12.2$ as an example, Fig.~\ref{x12} shows the scaled intensity
as a function of $q$. As for the $X=6.8$ microemulsions,
$I(q)/\phi $ increases initially with increasing concentration,
but now the region of $q^{-1}$ scattering is less pronounced.
Again, on varying the concentration some changes at large $q$ are observed,
suggesting that structural changes occur at the droplet level.

\begin{figure}[ht]
\centerline{\psfig{figure=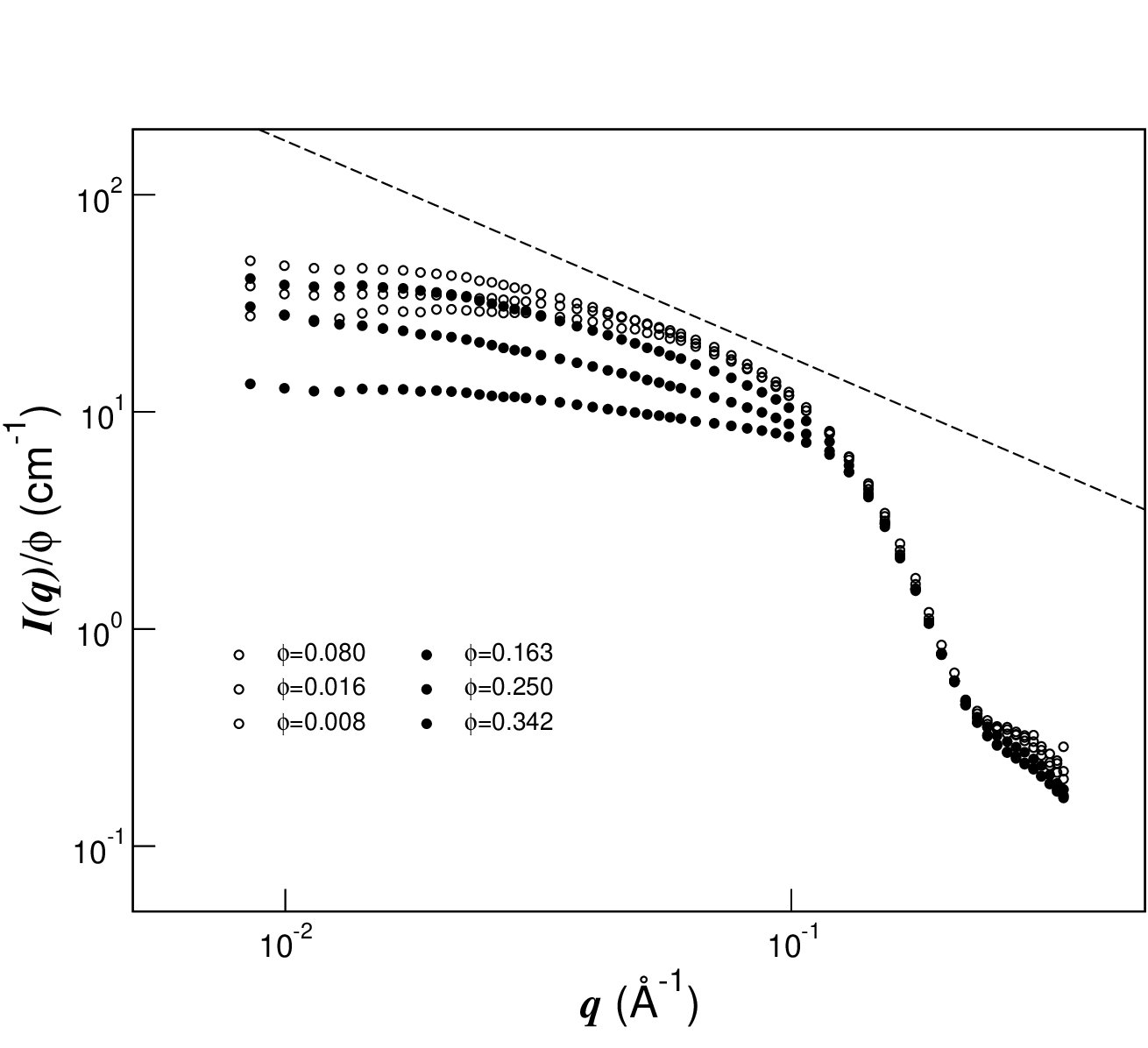,width=12 cm}}
\caption{Intensity scaled on the droplet volume fraction $\phi $
as a function of $q$ for different $\phi $ at $X=6.75$ and
[TBAC]/[AOT]=0.05.
The data have been divided in two groups with open circles denoting,
in order of increasing $I(q)/\phi $ along the left-hand side,
$\phi =$0.008, 0.016, 0.080, and
filled circles denoting, in order of decreasing $I(q)/\phi $
along the left-hand side, $\phi =$0.163, 0.250, and 0.342.
The dashed line shows $I(q)\propto q^{-1}$.
}
\label{tbac}
\end{figure}

Aside from changing the composition via the molar ratio
to move away from the peak in viscosity,
one can introduce TBAC to suppress the viscosity enhancement
altogether. In Fig.~\ref{tbac}, TBAC has been introduced at a
concentration that nearly eliminates
the maximum in viscosity in Fig.~\ref{salt}, [TBAC]/[AOT]=0.05.
The same initial increasing trend in $I(q)/\phi $ with increasing
concentration is observed as is the ensuing decrease
beyond $\phi \approx 0.07$.
Some power-law scattering is observed
over limited ranges of $q$
at the higher volume fractions, but
a region of $q^{-1}$ scattering is not present.
Also, the scattering from samples at $X=3$ (data not shown) did not
contain any power-law scattering at all.
It follows that we can conclude that for microemulsions at the
viscosity maximum the scattering intensity increases strongly
initially with concentration and
eventually presents a clear-cut
region where $I(q)\propto q^{-1}$. This behavior
points to a reorganization of aggregates with
increasing concentration into elongated structures.

\vskip0pt plus2mm
\subsection{Analysis of scattering results}

Extracting precise structural information from the scattering data requires
quantitative modeling, usually done by model fitting \cite{Ped97}.
In many cases model-independent
analyses can also be used very effectively
\cite{Gla80,War90,Ped95,Sch96b,Fri06}.
In this work we have only pursued model fitting
to a limited extent because of the lack of an appropriate
structure factor model that captures the coupling between positional and
orientational correlations in systems with
a significant degree of particle form anisotropy over an extended range of concentration.

\begin{figure}[ht]
\centerline{\psfig{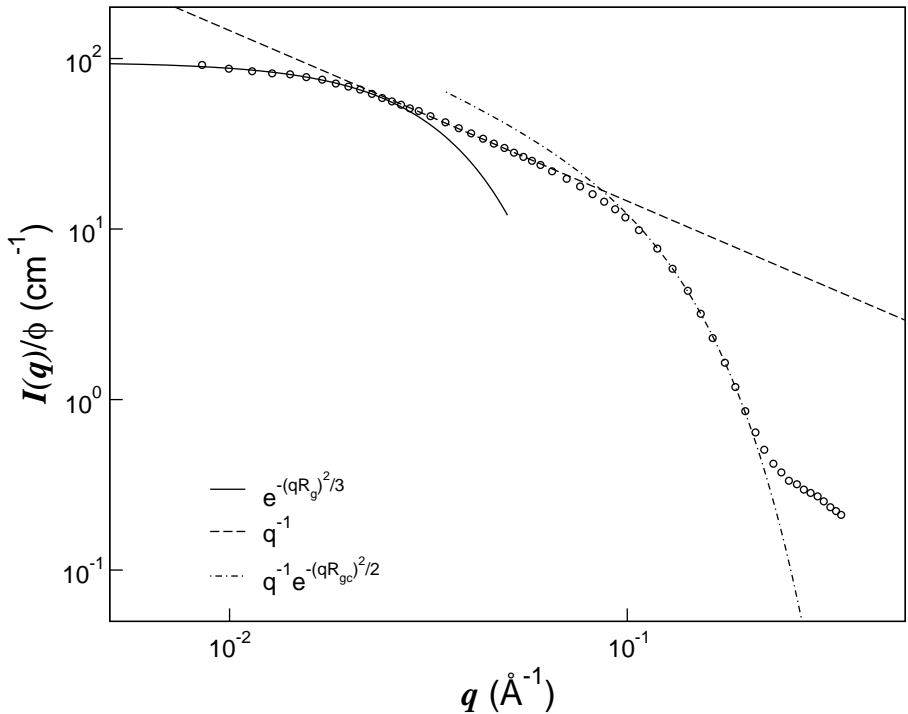}}
\caption{Intensity scaled on the droplet volume fraction $\phi $
as a function of $q$ for $\phi =0.07$ at $X=6.8$.
The curves illustrate three different asymptotic regimes as discussed
in the text.
}
\label{asymp}
\end{figure}

Since the instrumental
SAXS setup delivers data that extend almost two orders of magnitude in $q$,
it is worthwhile to begin with an asymptotic analysis.
In Fig.~\ref{asymp} we illustrate three
regimes of $q$ which seem to accommodate such an analysis.
At small $q$ the scattering is only sensitive to the overall
dimension of the scattering particles and we expect
from Guinier's law that the intensity should behave
as $I(q)\propto e^{-(qR_g)^2/3}$,
where $R_g$ is the radius of gyration. For higher
concentrations we define an apparent correlation
length $\xi $, viz. $I(q)\propto e^{-(q\xi )^2}$ \cite{Doi86}, which
crosses over to $R_g/\sqrt{3}$ under dilute conditions.
For cylindrical particles of length $L$
and cross-sectional radius $R$ with $L>R$, the Guinier
regime in principle requires $q\ll L^{-1}$.
At intermediate $q$ we observe that the intensity
varies inversely with $q$, pointing to cylindrical aggregates.
We dismiss the possibility that this power-law scattering comes from
highly random, fractal aggregates of spherical droplets
simply because the microemulsions here are equilibrium systems;
such fractal structures in dispersions of colloidal particles result
from irreversible, non-equilibrium aggregation processes \cite{Poo97}.
For cylindrical structures in the small-$q$, Guinier regime
we expect the following to hold in the dilute limit
\begin{equation}
I(q)/\phi \approx \Delta \varrho ^2 (\pi R^2L)
e^{-(qR_g)^2/3}
\label{guin}
\end{equation}
where the radius of gyration,
$R_g=\left(L^2/12+R^2/2\right)^{1/2}$,
and $\Delta \varrho $
is the electron density difference between the homogeneous particle
and the solvent. It follows that for sufficiently
large aspect ratios $L/R$ both $I(0)/\phi $ and $R_g$ are
linearly proportional to the cylinder length.
At larger $q$ the scattering no longer depends on the
actual cylinder length and provided that $L\gg R$
one might hope to find a regime $L^{-1}\ll q \ll R^{-1}$
within which \cite{Por82,Lin87,Sch91,Lon94}
\begin{equation}
I(q)\propto q^{-1}e^{-(qR_{gc})^2/2}
\label{porod}
\end{equation}
where $R_{gc}$ is a radius of gyration of the cross-sectional area.
Indeed, in the scattering data for $X=6.8$ in Fig.~\ref{x6}
we find a small linear regime on graphing $\ln (qI(q))$ versus $q^2$
and determine $R_{gc}\approx 11.4$\AA.
For a circular droplet cross-section
this corresponds to a radius of about 16.1 \AA.
A similar analysis for varying $\phi $ at this $X$ revealed
no systematic variation in $R$,
suggesting that droplets elongate without changing
their cross-section significantly.
More detailed analyses
of the cross-sectional structure of surfactant micelles
have been done using small-angle neutron scattering measurements
combined with contrast variation \cite{Sch96b}.
One slight worry is that Eq.~\ref{porod} is not able to capture
the rather abrupt cross-over in Fig.~\ref{asymp}
to a $q^{-1}$ dependence, but the
analysis here is strictly speaking
restricted to homogeneous, monodisperse
cylinders under neglect of concentration effects.

\begin{figure}[ht]
\centerline{\psfig{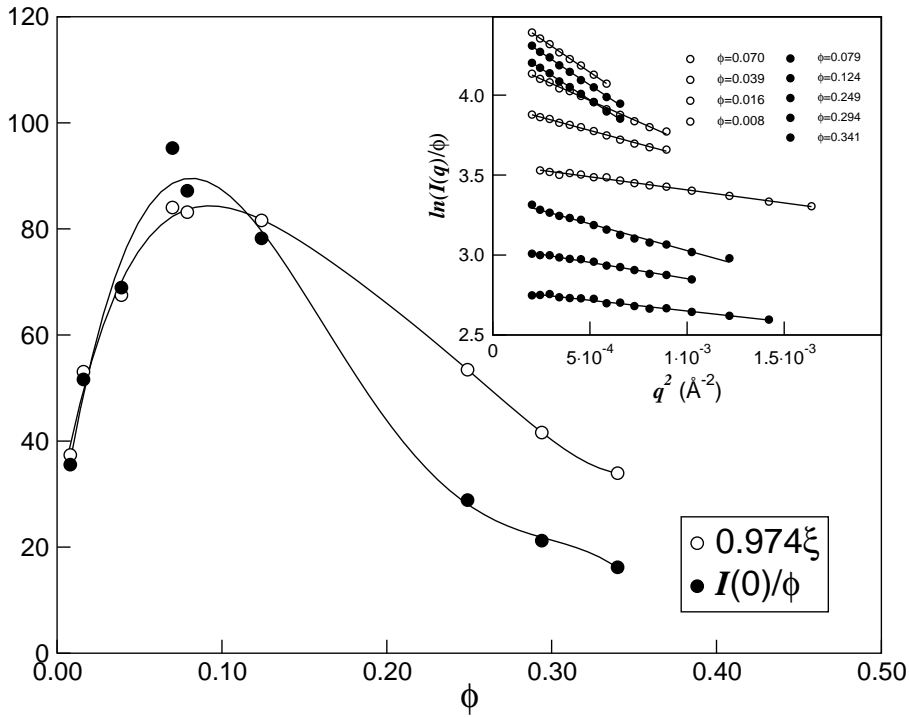}}
\caption{Comparison of the $q\to 0$ limit of $I(q)/\phi $ (cm$^{-1}$)
and the correlation length (\AA ), adjusted by
an arbitrary multiplicative factor,
as a function of droplet volume fraction for $X=6.8$.
Curves are drawn as guides to the eye.
The inset shows
the low-$q$ part of the data in Fig.~\ref{x6}
graphed for a Guinier analysis.
}
\label{guin1}
\end{figure}

\begin{figure}[ht]
\centerline{\psfig{figure=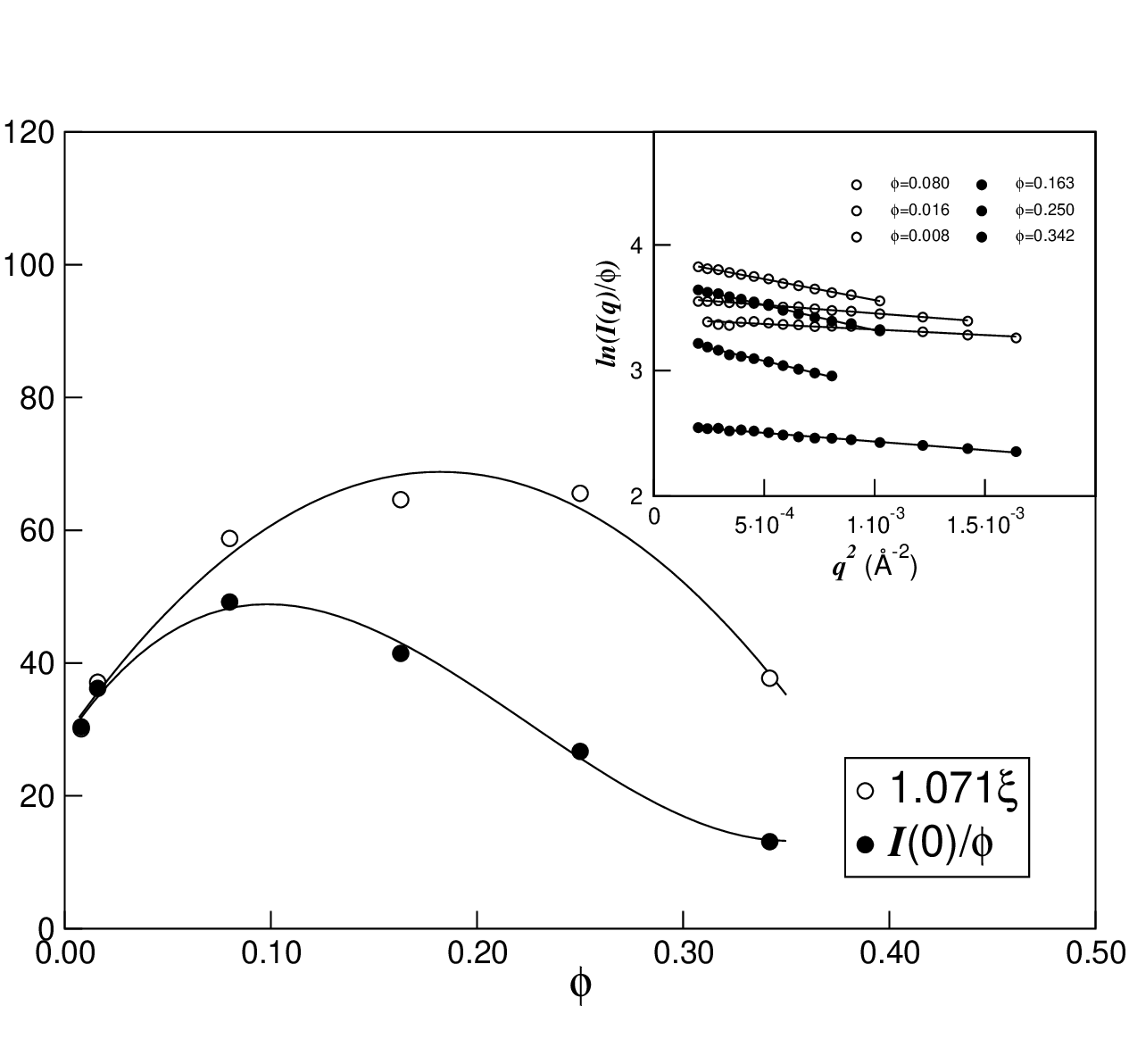,width=12 cm}}
\caption{Comparison of the $q\to 0$ limit of $I(q)/\phi $ (cm$^{-1}$)
and the correlation length (\AA ), adjusted by an arbitrary multiplicative factor,
as a function of droplet volume fraction for X=6.75 and [TBAC]/[AOT]=0.05.
Curves are drawn as guides to the eye. The inset shows
the low-$q$ part of the data in Fig.~\ref{tbac}
graphed for a Guinier analysis.
}
\label{guin2}
\end{figure}

As pointed out in connection with Eq.~\ref{guin},
for systems that are not too concentrated we expect
both $I(0)/\phi $ and $R_g$ to be roughly proportional
to the cylinder length provided the axial ratio $L/R$ is
significant. To examine whether this holds for these
microemulsions, we show in Fig.~\ref{guin1} the results
of extracting $I(0)/\phi $ and
$\xi $ from a Guinier analysis of the $X=6.8$
data in Fig.~\ref{x6}, i.e. as the intercept and slope
on a $\ln I(q)$ vs. $q^2$ graph of low-$q$-limiting data.
For comparative purposes
the correlation length in Fig.~\ref{guin1} has been shifted
by a multiplicative factor. One observes that as a function of $\phi $ both
$I(0)/\phi $ and $\xi $ pass through a maximum at $\phi ^*\approx 0.07$.
In addition, the fact that
they track each other closely for volume fractions below $\phi ^*$
suggests first of all that $\xi \approx R_g$ for this range
of concentrations; secondly, it points to
the low-$q$ scattering being caused by droplet
growth into elongated, perhaps cylindrical aggregates,
which proceeds by increasing $L$
at roughly constant $R$.
On adding the salt TBAC to the microemulsions, the same analysis
yields the results in Fig.~\ref{guin2}.
The intensity and correlation length are both reduced
for most volume fractions relative to the corresponding
values for the salt-free microemulsions at $X=6.8$.
Although maxima are observed again in both
$I(0)/\phi $ and $\xi $ as functions of $\phi $, the location
of the maximum in correlation length is shifted
to larger $\phi $ compared to in the absence of salt.
In addition, except for the two most dilute
microemulsions, $I(0)/\phi $ and $\xi $ do not exhibit the
same volume fraction dependence, suggesting very limited
cylindrical growth and small aspect ratios on the whole.

The behavior in Figs.~\ref{guin1} and \ref{guin2} is qualitatively
the same as for polymer-like reverse micelles, though the limited
growth here is in no way comparable.
Similarly to the microemulsions in this study, however,
the growth in those systems is
triggered by addition of water \cite{Lui90,Sch91,Eas92b}
and an extreme viscosity maximum
appears as well as a function of molar ratio \cite{Lui90}.
Similar maxima as in Fig.~\ref{guin1} have been interpreted as
reflecting the crossover from dilute to semi-dilute concentration
regimes \cite{Sch96}. For rigid cylinders the volume fraction
at which their rotational motion
starts to become impeded due to concentration effects
is $\phi ^*_1 \approx 6(R/L)^2$,
but static properties are not expected to be too seriously
affected until excluded-volume effects become
significant at another volume fraction,
$\phi ^*_2 \approx \frac{\pi}{2}(R/L)$ \cite{Doi86}.
With $\phi ^*\approx 0.07$ from Fig.~\ref{guin1},
regardless of whether $\phi ^*_1$ or $\phi ^*_2$ is assigned
as the overlap concentration,
we can expect axial ratios of ${\cal O}(10)$ close to
$\phi ^*$ for microemulsions with $X=6.8$.
In other words, it is indeed a rather modest degree of
cylindrical growth.

To extract a more precise measure of the initial
droplet growth, we attempt
to model some of the data quantitatively.
Anticipating droplet axial ratios of ${\cal O}(10)$, this is a difficult task
because orientational correlations are expected to have to be taken into account.
Models that incorporate orientational
correlations are available for
slender rigid rods \cite{Shi88,Doi88,vdS90,vdS92,Mae91},
but they are in severe disagreement with the experimental data
in Figs.~\ref{x6}-\ref{tbac}
at large $q$ due to their tending to the form factor of
an infinitely thin rod.
To remedy this situation, we follow the procedure
suggested by Weyerich et al. \cite{Wey92}
for approximately accounting for finite-thickness effects.
The scattering intensity is expressed as
\begin{equation}
I(q) = nP(q)\frac{S(q)}{F(q)} 
\label{intens}
\end{equation}
where $n$ is the number density and
$S(q)$ is the structure factor for slender hard rods
with $F(q)=S(q,n\to 0)$ as the slender rod form factor.
The dilute-limiting behavior is then $I(q)\approx nP(q)$, where
we take $P(q)$ as the orientationally-averaged form factor
of core-shell-structured cylinders \cite{Liv87},
\begin{eqnarray}
P(q) &=& \left(\frac{4\pi }{q^2}\right)^2\int_0^{\pi /2} d\beta \sin \beta
\left(\frac{\sin (qh\cos \beta )}{\sin \beta \cos \beta }\right)^2
\nonumber \\ & & \times
\int_0^\infty dR Rf(R)\left[
\varrho _{12}p_1J_1(qp_1R\sin \beta )+\varrho _{23}p_2J_1(qp_2R\sin \beta )
\right]^2
\label{formf}
\end{eqnarray}
where the inner and outer cylinder radii are given by
$p_1R$ and $p_2R$,
respectively, both proportional to the total cylinder cross-sectional
radius $R$. In addition, $h=L/2$,
$J_1(x)$ is the first-order Bessel function of the first kind,
$\varrho _{12}=\varrho _1-\varrho _2$ is the electron density difference
between the core and shell of the cylinder,
and $\varrho _{23}=\varrho _2-\varrho _3$ is
the electron density difference between the shell
and the solvent.
In Eq.~\ref{formf} the cylinder radius $R$ is taken as polydisperse.
We assign a log-normal distribution \cite{Pus82},
$f(R)$, characterized by a mean cross-sectional radius ${\bar R}$
and a standard deviation $\sigma _R$, to govern the polydisperse
radius. For simplicity, the core and core-plus-shell radii
are assumed proportional to the total radius so
that only a single polydispersity index, $\sigma _R/{\bar R}$,
is needed.

\begin{figure}[ht]
\centerline{\psfig{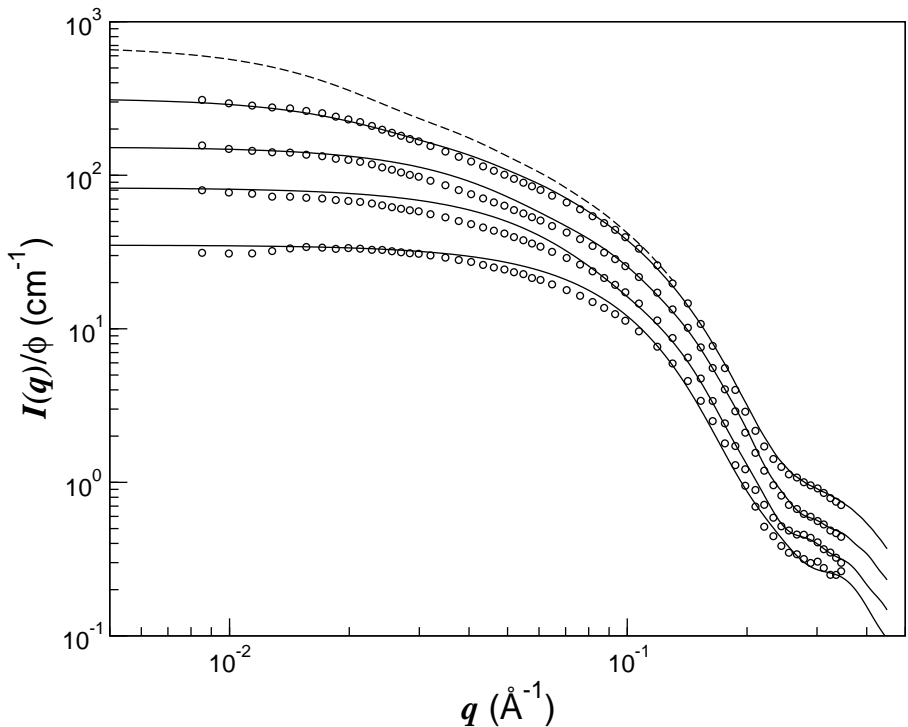}}
\caption{
Intensity scaled on the droplet volume fraction $\phi $
as a function of $q$ for different $\phi $ at $X=6.8$,
in order of increasing $I(q)/\phi $ along the left-hand side,
$\phi =$0.008, 0.016, 0.039, and 0.070.
For clarity the data have been offset vertically
by $1.5$, $1.5^2$, and $1.5^3$,
relative to scattering curve for  $\phi =$0.008.
The solid curves are fits of Eqs.~\ref{intens}-\ref{vds} to
the data. Shown also, as the dashed curve,
is the form factor in Eq.~\ref{formf} for $\phi =$0.070.
}
\label{fit}
\end{figure}

Inspection of the SAXS data in Figs.~\ref{x6}-\ref{tbac}
shows no evidence of a structure factor peak in the intensity.
Instead, beyond the overlap concentration
the scattering at small $q$ is increasingly suppressed as the
concentration is increased. At still higher concentrations, peaks in the
intensity do appear \cite{Ber95}.
This behavior is qualitatively consistent with the expected
behavior of slender hard rods under relatively
dilute conditions \cite{vdS90,vdS92}.
Guided by this observation we have used the structure factor of
slender hard rods, derived by van der Schoot and Odijk \cite{vdS90},
which is given by
\begin{eqnarray}
S(q) &=& F^2(q)/\left[F(q)+2cF^2(q)-5cG(q)/4\right]
\label{vds}
\end{eqnarray}
where $c=\pi nRL^2/2$, $F(q)=2Si(qL)/qL-(\sin (qL/2)/(qL/2))^2$
is the form factor of a slender
rod, given in terms of the sine integral $Si(x)$,
and $G(q)=3(qL)^{-2}(1-\sin (qL)/qL)-F(q)/2$.
Polydispersity in cylinder length is neglected because,
as observed by Weyerich et al.
for unimodal distributions \cite{Wey92},
the polydisperse slender hard-rod structure factor \cite{vdS92}
is insensitive to polydispersity in length.
In keeping with the limitations of using the slender hard-rod
structure factor of van der Schoot and Odijk, we restrict attention to
concentrations up to the overlap concentration,
$\phi \le \phi ^*\approx 0.07$ for $X=6.8$, where we
see the largest droplet growth. In Fig.~\ref{fit} these data
are shown along with fits of Eqs.~\ref{intens}-\ref{vds}.
In the fits we have constrained the
parameters so that
$\phi _w=n\pi p_1^2\langle R^2\rangle L$ and
$\phi = n\pi \langle R^2\rangle L$ are fulfilled,
with $\phi _w$ and $\phi $ the volume fractions of water and
water plus surfactant, respectively. Furthermore,
the cylinder radius that enters
the monodisperse hard-rod structure factor has been equated with the
average radius ${\bar R}$ used in $P(q)$.
The necessary integrations in Eq.~\ref{formf} were completed numerically
using Simpson's rule.

The analysis is complicated by the
change in basic structure of the droplets. Nevertheless,
as shown in Fig.~\ref{fit}, model fits can be produced
that are in reasonable accord with the experimental data.
The results are consistent with the asymptotic analysis
and reflect a significant elongation with increasing
droplet volume fraction. In going from the most dilute to
the most concentrated
microemulsion in Fig.~\ref{fit}, the cylinder length increases
from 45 up to 270 \AA\, with very slight changes in the cross-sectional
parameters. This growth corresponds to axial ratios $L/R$ increasing from
2.8 to 19.2. As seen in Fig.~\ref{fit}, the agreement between model
fits and data improves with increasing axial ratio, which we
attribute to the use of a slender-rod structure factor. Though structure
factor effects are completely negligible at the lowest volume
fraction in Fig.~\ref{fit},
the analysis suggests that they are quite significant at higher
volume fractions. To illustrate the extent of the effect of the
structure factor, we show in Fig.~\ref{fit} essentially what amounts
to the form factor, $nP(q)/\phi $, for the most concentrated sample.
The ratio
between the model fit and this quantity is $S(q)/F(q)$ in Eq.~\ref{intens}
at this concentration. It is worth noting that the structure
factor matters up to fairly large values of $q$.

The asymptotic analysis produced a cylinder thickness of about 16.5 \AA,
but in the model fits of Fig.~\ref{fit}
smaller values in the range 11.5-11.9 \AA~were used.
We attribute this discrepancy to the implicit use
of a homogeneous rather than a core-shell cylinder model in
the asymptotic analysis. As a consequence, $R_{gc}$ is an
apparent value weighted by the cross-sectional electron density distribution
in the droplet \cite{Lon94}.
Some additional comments regarding the fits are in order.
The model fits in Fig.~\ref{fit} correspond to the
core-shell cylinder model in Eq.~\ref{formf}
with contrast factors $\varrho _{12}<0$ and $\varrho _{23}>0$.
Comparison with results of similar analyses done at larger molar ratios
\cite{Rob89,Arl01}, $X\approx 40$, shows that the fits here require
a broader interfacial region in terms of the
shell thickness, $(p_2-p_1){\bar R}$. At larger $X$ this could
be set to a small value, based on a large-$q$, Porod analysis \cite{Rob88},
with a large $\varrho _2$ value of 850 $e$/nm$^3$.
In contrast, the model fits here are generated with a 5 \AA\, shell
thickness, close instead to that deduced from small-angle
neutron scattering on AOT micelles \cite{Kot85} and used
by others \cite{Hil90}, and a correspondingly weaker contrast of the shell,
$\varrho _2\approx $ 440-480 $e$/nm$^3$.
Similarly broad interfaces result also
on changing the cylindrical core-shell form factor
in Eq.~\ref{formf} to other basic core-shell structures,
such as spheres or prolate ellipsoids. The quality of the fits
at the lowest $\phi $ in Fig.~\ref{fit} did not improve with
such a change in droplet structure.

Eicke and Rehak \cite{Eic76} and Maitra \cite{Mai84},
based on a spherical droplet model, determined an area per
surfactant headgroup that was found to decrease
dramatically with decreasing molar ratio,
a trend that seems to extrapolate
to a $X=0$ value far below that obtained from small-angle neutron scattering
measurements on
micelles \cite{Kot85}. Furthermore, Maitra suggested a model
for AOT microemulsions at low $X$ values
in which the surfactant-water interface is overcrowded by surfactant,
effectively reducing the area per headgroup.
This was speculated to give rise to an increased attraction
that in turn caused the viscosity enhancement \cite{Ber95}.
Considering changes in droplet structure instead, we
determine the area per headgroup as $2\pi p_1{\bar R}L(n/n_{\rm surf})$,
which yields values ranging from 49 \AA$^2$ at the lowest $\phi $
to 56 \AA$^2$ at the highest $\phi $ in Fig.~\ref{fit}.
This range of values is more in line with scattering results on
AOT micelles \cite{Kot85} and microemulsions at larger $X$
\cite{Kot84,Rob89,Arl01}. In other words, the trend of a
decreasing area per headgroup with decreasing molar
ratio \cite{Mai84,Eic76} is likely a
consequence of constraining the droplets to be spherical in shape.
In a similar way, it is possible that
a change in aggregate structure from globular to
cylindrical can also explain irregular trends
in the apparent molar volume of solubilized water, as
conceived by Yoshimura et al. \cite{Yos00}.

Measurements at similar molar ratios to the ones studied here
have been carried out on related AOT microemulsions where the
usual Na$^+$ counterion has been exchanged
for a variety of monovalent and divalent cations
\cite{Pet91,Eas92,Eas93a,Eas93b,Eas94}.
As in the present measurements, power-law, $q^{-1}$ scattering has
been used to infer elongated, rodlike droplets in many of these
systems and the tendency to form cylindrical droplets has been
correlated with the
hydrated volume of the counterion \cite{Eas93a,Eas93b}.
Moreover, it is emphasized that the usual AOT system, from
the Na-salt, forms only spherical aggregates in the L$_2$ phase
regardless of molar ratio
along with a few other AOT systems, such as
AOT with Ca$^{2+}$ counterions.
However, recently Pan and Bhatia \cite{Pan08} found a viscosity maximum
for AOT microemulsions with Ca$^{2+}$ counterions
at somewhat larger values of $X$ than those in Fig.~\ref{salt}.
If there are composition
ranges within which droplets based on the usual AOT form
cylindrical structures then one is led to believe that
such composition ranges also exist for AOT microemulsions
with, for instance, Ca$^{2+}$ counterions.
Indeed, Pitzalis et al. \cite{Pit00} conclude,
based on NMR self-diffusion measurements, that non-spherical
shapes are rather the rule than the exception
at low molar ratios for
AOT microemulsions with Na$^+$ and Ca$^{2+}$ counterions.
In addition, nuclear spin relaxation measurements, done at somewhat larger
molar ratios than the viscosity maximum, seem to extrapolate to sizeable
shape anisotropies at the molar ratios of the viscosity maximum
\cite{Car89}.
With scattering data as function of
both molar ratio and droplet concentration in hand
(see Figs.~\ref{x6}-\ref{tbac})
one sees that the $q^{-1}$ scattering regime is
easily missed, since it only appears clearly
at certain volume fractions.

\vskip0pt plus2mm
\section{Conclusions}

Viscosity measurements have been conducted on AOT/H$_2$O/decane
microemulsions confirming a maximum in the viscosity as a function of
H$_2$O/AOT molar ratio. Exchange of the continuous oil phase for
carbon tetrachloride and sufficient addition of salt
eliminates the viscosity maximum.
Systematic small-angle X-ray scattering measurements
have been conducted and analyzed. They reveal that these
microemulsions produce globular aggregates at very low molar ratios.
At low-to-intermediate molar ratios they yield
somewhat anisotropic, elongated structures, presumably due to incomplete
hydration of the headgroup and packing constraints imposed, e.g.,
by headgroup interactions and oil penetration of surfactant tails.
In addition, the structures grow in length
with increased concentration of water and surfactant,
as anticipated already by Wennerstr\"om and Lindman \cite{Wen79}.
At still larger molar ratios less elongated droplets
are produced. The viscosity of the microemulsions appears
to reflect directly, albeit qualitatively, the degree
of structural anisotropy in this region of molar ratio.
The results suggest that the AOT molecule, regardless
of the identity of the counterion, preferentially packs
into non-spherical aggregates at lower molar ratios.
However, the precise composition range where this occurs and the
degree of structural anisotropy is no doubt
a function of the identity of the counterion.

\vskip0pt plus2mm
\subsection*{Acknowledgment}

Financial support by the Swedish Research Council
is gratefully acknowledged.

\end{document}